\documentclass{article}
\usepackage{spconf,amsmath,graphicx}
\usepackage{csquotes}
\usepackage{tikz}
\usepackage{pgfplots}
\usepackage{url}
\usepackage{todonotes}
\usepackage{xcolor}
\usepackage{amssymb}
\usepackage{hyperref}
\hypersetup{
    colorlinks=true,
    linkcolor=red,
    filecolor=magenta,      
    urlcolor=magenta,
}

\usetikzlibrary{patterns}
\mathchardef\mhyphen="2D

\title{Investigating Deep Neural Structures and their Interpretability in the Domain of Voice Conversion}
\name{Samuel J. Broughton, Md Asif Jalal, Roger K. Moore}
\address{Dept. Computer Science, University of Sheffield, UK}
\begin{document}
%\ninept
%
\maketitle
\begin{abstract}
Generative Adversarial Networks (GANs) are machine learning networks based around creating synthetic data. Voice Conversion (VC) is a subset of voice translation that involves translating the paralinguistic features of a source speaker to a target speaker while preserving the linguistic information. The aim of non-parallel conditional GANs for VC is to translate an acoustic speech feature sequence from one domain to another without the use of paired data. In the study reported here, we investigated the interpretability of state-of-the-art implementations of non-parallel GANs in the domain of VC. We show that the learned representations in the repeating layers of a particular GAN architecture remain close to their original random initialised parameters, demonstrating that it is the number of repeating layers that is more responsible for the quality of the output. We also analysed the learned representations of a model trained on one particular dataset when used during transfer learning on another dataset.  This showed extremely high levels of similarity across the entire network. Together, these results provide new insight into how the learned representations of deep generative networks change during learning and the importance in the number of layers.\footnote{Audio samples available at: \texttt{\url{https://samuelbroughton.github.io/interpretability-demo-2020}}.}
\end{abstract}


\section{Introduction}

Deep Learning networks have been shown to exhibit superior abilities in a range of problem domains \cite{Wang, Zhu2017, Kaneko2019cycle}. However, such networks are \emph{black-box representations} in terms of their interpretability \cite{Zhang2018}, and this can mitigate against informed decision making when selecting appropriate network configurations.

One problem domain, voice conversion (VC), or voice style transfer, is a technique aimed at modifying the linguistic style of speech while preserving the linguistic information contained therein \cite{Mohammadi2017, Toda2016, Lorenzo-Trueba2018}. VC can be formulated as a regression problem with the aim of building a function in which the features of a source speaker A can be mapped to a target speaker B \cite{Stylianou1998, Toda2016, Lorenzo-Trueba2018}. Applications of VC include modifying speaker identity in text-to-speech (TTS) systems \cite{Kawanami2003}, aiding those with vocal disabilities \cite{Nakamura2012} and generating accents for assisted language conversion in domains such as real-time language translation and device-assisted language learning \cite{Felps2009}. 

Historically, methods employed to achieve VC have included mapping code books \cite{Abe1990}, Gaussian mixture models (GMMs) \cite{Stylianou1998, Kawanami2003, Toda2007} and artificial neural networks (ANNs) \cite{Desai2009, Mohammadi2014}. However, variations of generative adversarial networks (GANs) \cite{Goodfellow} have recently shown success in a range of different domains, such as producing convincingly real images and videos \cite{Zhu2017, KarrasNVIDIA}, enhancing the quality of images \cite{Wang}, generating new music \cite{Dong2018} and, of interest here, a methodology for achieving VC \cite{Hsu, Kaneko2017, Kaneko2019star, Kaneko2019cycle, Paul2019, Lee}.

Some of the VC methods mentioned above can be categorized as either parallel or non-parallel. Parallel VC refers to source and target speaker utterances being perfectly aligned \cite{Toda2016, Lorenzo-Trueba2018}. Such data can be a laborious task to collect. Furthermore, once collected, the data would need to be pre-processed with automatic time alignment which can fail, resulting in other methods of correction. However, GANs are able to learn mapping functions between data of similar domains and so mitigate the need for a parallel dataset \cite{Zhu2017}.  Recent state-of-the-art non-parallel generative VC architectures include CycleGAN-VC2 \cite{Kaneko2019cycle} and StarGAN-VC2 \cite{Kaneko2019star}. Both make use of a gated convolutional neural network (CNN) \cite{Dauphin2017}, identity-mapping loss \cite{Taigman} and \textit{2-1-2D CNN} architecture \cite{Kaneko2019cycle}.

A major advantage of using the StarGAN \cite{Choi2018} framework when compared to CycleGAN \cite{Zhu2017}, is the ability to perform multi-domain conversion whilst only requiring a single generator. With regards to VC, the StarGAN framework allows for learned mapping functions between multiple speakers. Extending this framework, StarGAN-VC \cite{Kameoka2018} and StarGAN-VC2 \cite{Kaneko2019star} make various modifications including updates to the training objective and alterations to the network architecture. 

However, despite StarGAN-VC2 demonstrating superior VC in both objective and subjective experiments when compared to StarGAN-VC \cite{Kaneko2019star}, there has been very little investigation into the interpretability of it's network representations - as is the case with many deep multi-layer neural networks, especially GANs \cite{voynov2019rpgan}. Being able understand how interpret deep multi-layer GANs would benefit the development of new generative techniques and the improve the efficiency of current methods, as interpretability studies have began to do so with discriminative models \cite{mahendran2015understanding, Raghu2017}. The motivation for this paper is to provide some insight into the underlying generation process by focusing on the learned network representations and network depth.

%MAJ: "We" again. This is strictly prohibited 
% randomly initialised states

In this work, we conducted an evaluation of learned network representations by performing Singular Vector Canonical Correlation Analysis (SVCCA) \cite{Raghu2017} in a range of different experiments using an adaptation of the StarGAN-VC2 network. The aim was to provide insight into the interpretability of GANs for VC by addressing the similarity of optimally trained networks and their random initial states. This was achieved by conducting experiments with networks including frozen layers, observing how quickly networks reached their optimal representations, exploring the effects of modifying the size of networks and investigating learned network representations when trained using transfer learning. 

% analysing the conversion when network size is modified (not explore)

The rest of the paper is structured as follows: Section \ref{sec:net} outlines the generative network architecture used, Section \ref{sec:svcca} discusses SVCCA and the motivation to use it in this work, Section \ref{sec:experiments} describes the research questions and experimental conditions, Section \ref{sec:results} discusses the results and probable implications, and Section \ref{sec:conclusion} draws the conclusion.

\section{Generative Network Architecture} \label{sec:net}

% we are using the network presented by cite 
% more interesting than just state of art - is deep 

The network architecture implemented for the experiments presented in this paper was based on StarGAN-VC2 \cite{Kaneko2019star}, which allows for non-parallel many-to-many learned mappings for VC.

\subsection{Training objectives}

The main objective of the StarGAN framework \cite{Choi2018} is to learn many-to-many mapping functions between multiple domains whilst only using a single generator $G$. StarGAN does this by conditioning itself on `one-hot' representations of domain codes $c \in \{1, ..., N\}$, where $c$ and $N$ indicate the domain code the number of domains, respectively. More specifically in StarGAN-VC2, $G$ can be formulated as the mapping function $G(x, \hat{c}) \xrightarrow{} \hat{x}$, taking an acoustic input feature sequence $x \in \mathbb{R}^{Q \times T}$ and target domain code $\hat{c}$ to generate an acoustic output feature sequence $\hat{x}$. StarGAN-VC2 does this by making use of an adversarial loss \cite{Goodfellow}, reconstruction or cycle-consistency loss \cite{Zhu2017} and identity-mapping loss \cite{Taigman}.

\textbf{Adversarial loss} is used in GANs to encourage generated data, conditioned on target domain code, to be indistinguishable to that of real data \cite{Choi2018}:

\begin{align}
    \mathcal{L}_{adv} &= \mathbb{E}_{(x, c) \sim P(x, c)} [\log D(x, c)] \nonumber \\
    &+ \mathbb{E}_{x \sim P(x), \hat{c} \sim P(\hat{c})} [\log(1 - D(G(x, \hat{c}), \hat{c}))],
    \label{eq:s-t_adv_loss}
\end{align}

where $D$ is a real/fake discriminator that attempts to maximise this loss to learn the decision boundary between real and fake features. $G$ attempts to minimize this loss by generating an output indistinguishable to the real acoustic features of target domain $\hat{c}$. 

As discussed in the StarGAN-VC2 study \cite{Kaneko2019star}, when handling both hard negative and easy negative samples (e.g. same speaker domain conversion and different speaker domain conversion), this condition can make it difficult to bring generated output data close to real target data. Therefore, \textit{source-and-target} conditional adversarial loss \cite{Kaneko2019star} is used to help $G$ generate an output closer to the real target data. However, during pre-experiments we found that only using target domain input in $G$ and using both source-and-target domain inputs in $D$ yielded a better output quality without affecting speaker similarity. The modified source-and-target adversarial objective is defined as:

\begin{align}
    \mathcal{L}_{st \mhyphen adv} &= \mathbb{E}_{(x, c) \sim P(x, c), \hat{c} \sim P(\hat{c})} [\log D(x, \hat{c}, c)] \nonumber \\
    &+ \mathbb{E}_{(x, c) \sim P(x, c), \hat{c} \sim P(\hat{c})} [\log D(G(x, \hat{c}), c, \hat{c})],
    \label{eq:s-t_adv_loss}
\end{align}

\textbf{Cycle-consistency loss} is used in order to guarantee that the converted output feature sequence preserves the source characteristics of input feature sequence $x$ \cite{Zhu2017, Choi2018}:

\begin{align}
	\mathcal{L}_{cyc} = \mathbb{E}_{(x, c) \sim P(x, c), \hat{c} \sim P(\hat{c})}[||x-G(G(x, \hat{c}), c)||_{1}].
	\label{eq:cycle_loss}
\end{align}

This cyclic constraint encourages $G$ to reconstruct the original input feature $x$ from the generated output $\hat{x}$ and source domain code $c$. This helps $G$ to preserve the linguistic information of the speech \cite{Kameoka2018}.

\textbf{Identity-mapping loss} is employed to encourage the preservation of input feature identity within generated output data \cite{Taigman}:

\begin{align}
    \mathcal{L}_{id} &= \mathbb{E}_{(x, c) \sim P(x, c)}[||G(x, c) - x||].
    \label{eq:im_loss}
\end{align}

Identity-mapping loss has previously been used in image-to-image translation for colour preservation \cite{Zhu2017}.

The \textbf{full objective} can be summarised as follows:

\begin{align}
    \mathcal{L}_{G} &= \mathcal{L}_{st \mhyphen adv} + \lambda_{cyc} \mathcal{L}_{cyc} + \lambda_{id} \mathcal{L}_{id} \label{eq:vc2_full_G} , \\
    \mathcal{L}_{D} &= - \mathcal{L}_{st \mhyphen adv} \label{eq:vc2_full_D},
\end{align}

where $\lambda_{cyc}$ and $\lambda_{id}$ are hyperparameters for each term. Here, $G$ aims to minimise the loss whilst $D$ is trying to maximise it.

\subsection{Network architectures}

The fully convolutional GAN architecture used in the study reported here allows for acoustic input feature sequences of arbitrary sizes.

\textbf{Generator}: The input to $G$ was an image of size $Q \times T$ of an acoustic feature sequence $x$, where $Q$ and $T$ are the feature dimension and sequence length, respectively. A \textit{2-1-2D CNN} \cite{Kaneko2019cycle, Kaneko2019star} architecture was used to construct $G$. 2D convolutions are well suited for holding the original data structure whilst 1D convolutions work well at dynamically changing the data \cite{Kaneko2019cycle}. The implementation specifically used a gated CNN \cite{Dauphin2017}, which allowed for relevant features to be selected and propagated based on previous layer states. The effectiveness of a gated CNN for VC has already been confirmed in previous studies \cite{Kameoka2018, Kaneko2017}.

Conditional domain specific style code was injected in the 1D CNN architecture by a modulation-based method \cite{Kaneko2019star}. Conditional instance normalisation (CIN) \cite{Dumoulin2016, huang2017arbitrary} was used to modulate parameters in a domain-specific manner:

\begin{align}
    \text{CIN}(f, \hat{c}) = \gamma_{\hat{c}} (\frac{f - \mu (f)}{\sigma (f)}) + \beta_{\hat{c}},
    \label{eq:cin}
\end{align}

where $\mu(f)$ and $\sigma(f)$ are the average and standard deviation of feature $f$ and $\gamma_{\hat{c}}$ and $\beta_{\hat{c}}$ are domain-specific scale and bias parameters, respectively.

The 1D repeating blocks were not residual because the use of skip connections was reported to result in partial conversion \cite{He2016}.

\textbf{Real/Fake Discriminator}: A 2D gated CNN \cite{Dauphin2017} was used for the architecture of the real/fake discriminator $D$, which has been formulated as a projection discriminator \cite{Miyato}, as seen in StarGAN-VC2 \cite{Kaneko2019star}. $D$ outputs a sequence of probabilities, calculating how close the input acoustic feature sequence $x$ is to domain $c$. 

\begin{figure*}[ht]
    \centering
	\includegraphics[width=\linewidth]{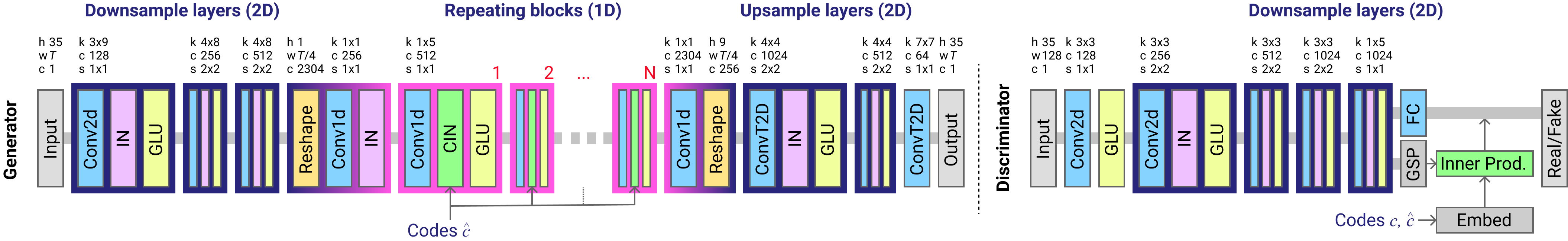}
	\caption{Network architectures of the fully convolutional \cite{Long} generator and discriminator based on StarGAN-VC2 \cite{Kaneko2019star}. In the input, output and reshape layers `h', `w' and `c' represent the height, width and channel number respectively. In the Conv2d, Conv1D and ConvT2D convolution layers, `k', `c' and `s' represent the kernel size, channel number and stride, respectively. `IN', `GLU', `GSP' and `FC' denote instance normalisation \cite{Ulyanov}, gated linear unit \cite{Dauphin2017}, global sum pooling and fully connected layers, respectively. $N = 9$ repeating 1D CNN blocks were used for all experiments unless otherwise specified.} 
	\label{fig:network}
\end{figure*}

\section{SVCCA on Deep Neural Representations} \label{sec:svcca}

Singular Vector Canonical Correlation Analysis (SVCCA) is an extension of Canonical Correlation Analysis (CCA), a method used in statistics to measure the similarity of two vectors formed by some \textit{underlying process} \cite{Hotelling1992, Hardoon2004, Morcos2018}. In the case of deep neural networks, these are the \enquote{neuron activation vectors} formed from training on a particular dataset \cite{Morcos2018, Raghu2017}. A single neuron activation vector is the output of a single neuron of a layer in a network. Combining the outputs of all neurons for a particular layer in a network results in a set of multidimensional output \cite{Morcos2018, Raghu2017}. Subsequently, CCA can be used to compare the similarity between two layers of the same network, similar networks using layers of same/differing dimensionality, or a given layer at different stages of training \cite{Morcos2018}.

SVCCA is an extension to CCA that involves a preprocessing step \cite{Morcos2018, Raghu2017}. The authors of \cite{Raghu2017} explain that SVCCA takes the same inputs as CCA, for example two layers of a neural network $l_{1}$ and $l_{2}$ that each contain a set of neuron activation vectors. SVCCA then factorises the vectors by computing Singular Value Decomposition (SVD) over each layer to obtain subspaces $l'_{1} \subset l_{1}$ and $l'_{2} \subset l_{2}$. These subspaces contain the most important variance directions, which can account for 99\% of the variance in input layers $l_{1}$ and $l_{2}$ \cite{Raghu2017}. CCA is then performed on $l'_{1}$ and $l'_{2}$ to return the correlation coefficients, providing a measure of similarity of the two layers.

The motivation for SVCCA in this paper is to provide a similarity metric for the comparison of various layers in the generator network. This allows for the interpretation of learned network representations at different stages of training. 

\section{Experiments} \label{sec:experiments}

\textbf{Datasets}: To evaluate our methods, we made use of the Device and Produced Speech Dataset \cite{Gautham2015}, as seen in the multi-speaker VC task in the Voice Conversion Challenge 2018 (VCC2018) \cite{Lorenzo-Trueba2018} and the English Multi-speaker Corpus for CSTR Voice Cloning Toolkit (VCTK) \cite{Veaux2017}. We used a subset of both datasets in all experiments except during transfer learning where the initial model was trained using the VCTK dataset.

In both datasets four speakers were selected covering all inter- and intra-gender conversions. In the VCTK dataset we selected speakers labelled \texttt{p229}, \texttt{p236}, \texttt{p232} and \texttt{p243}; speakers \texttt{p229} and \texttt{p236} are female, and speakers \texttt{p232} and \texttt{p243} are male. The data from VCC2018 mimicked the data used to test StarGAN-VC2 \cite{Kaneko2019star}, whereby \texttt{VCC2SF1} and \texttt{VCC2SF2} are female speakers, and \texttt{VCC2SM1} and \texttt{VCC2SM2} are male speakers. Speakerwise normalisation was conducted as a pre-process.

For each experiment, $4 \times 3 = 12$ source-and-target pair mappings were learnt for each single model trained on both datasets. All the recordings for both datasets were downsampled to 22.05 kHz. 36 Mel-cepstral coefficients (MCEPs) were extracted from each recording. The logarithmic fundamental frequency ($\log F_{0}$) and aperiodicities (APs) were extracted every 5 ms using the WORLD vocoder \cite{Morise2016}.

\textbf{Conversion process}: The conversion process mimicked that of StarGAN-VC \cite{Kameoka2018} and StarGAN-VC2 \cite{Kaneko2019star}, by not using any form of post filtering \cite{Kaneko2017filter, Kaneko2017specto} or powerful vocoding \cite{Oord2016wavenet, Tamamori2017vocoder} and just focusing on MCEP conversion\footnote{Audio samples available at: \texttt{\url{https://samuelbroughton.github.io/interpretability-demo-2020}}.}. As in previous studies, the WORLD vocoder \cite{Morise2016} was used to synthesise speech, directly taking APs and converting the $\log F_{0}$ using a logarithm Gaussian normalised transformation \cite{Liu}.

\textbf{Network implementations}: Figure \ref{fig:network} presents the network architectures for $G$ and $D$, influenced by StarGAN-VC2 \cite{Kaneko2019star} and CycleGAN-VC2 \cite{Kaneko2019cycle}. The networks were initially trained for $2 \times 10^{5}$ batch iterations on both datasets. During transfer learning, optimal models trained on the VCTK dataset were selected and trained for an extra $1 \times 10^{5}$ batch iterations on the VCC2018 dataset. During the training of the networks for all experiments, the states for $G$ and $D$ were saved at every $1 \times 10^{4}$ batch iterations. All networks were trained using the Adam optimizer \cite{Kingma} with a momentum term $\beta_{1}$ set to $0.5$. The batch size was set to $16$ and we randomly cropped segments of 512 frames from randomly selected sentences. Learning rates for $G$ and $D$ were both set to $0.0001$, $\lambda_{cyc} = 10$ and $\lambda_{id} = 5$. $\mathcal{L}_{id}$ was only used for the first $10^{4}$ iterations. $9$ repeating 1D CNN blocks were used for all experiments unless otherwise specified. Least squares GAN \cite{mao2017least} was used for a GAN objective.

\textbf{Experimental investigation}: Experiments were conducted in order to provide insights into questions relating to the interpretability of the trained networks.  \textbf{Experiment 1} addressed the issue as to how similar the learned representations of the optimally trained network are to its random initialisation.  \textbf{Experiment 2} addressed the question of how similar the learned representations of networks trained via transfer learning on a new dataset are to their previously optimal representations when trained on the original dataset. \textbf{Experiment 3} addressed the issue as to how similar the learned representations of networks with various frozen repeating layers are.  \textbf{Experiment 4} addressed the question of how the quality of the output feature sequence changes with networks of a differing number of repeating layers.

\section{Results and Discussions} \label{sec:results}

\begin{figure}[t]
	\centering
	\begin{tikzpicture}
	\begin{axis}[
	width=\linewidth,
	height=6cm,
	bar width=0.15cm,
	enlarge x limits={abs=0.4cm},
	ylabel=CCA Distance,
	xlabel=Batch Iteration ($\times10^{5}$),
	xlabel style={name=xlabel},
	ymin=0.3, ymax=1,
	ytick={0.4, 0.5, 0.6, 0.7, 0.8, 0.9, 1},
	symbolic x coords={0.1, 0.2, 0.3, 0.4, 0.5, 0.6, 0.7, 0.8, 0.9, 1, 1.1, 1.2, 1.3, 1.4, 1.5},
	xtick={0.1, 0.3, 0.5, 0.7, 0.9, 1.1, 1.3, 1.5},
	legend style={at={(xlabel.south)}, yshift=-1ex, anchor=north},
	]
	\addplot[color={rgb,255:red,0; green,0; blue,255}, mark=*]  coordinates{
		(0.1, 0.926190543)
		(0.2, 0.87242139)
		(0.3, 0.828600539)
		(0.4, 0.791094201)
		(0.5, 0.754200016)
		(0.6, 0.72453968)
		(0.7, 0.700337336)
		(0.8, 0.680183892)
		(0.9, 0.662168058)
		(1, 0.646559865)
	};
	\addlegendentry{D-sampling layers}
	\addplot[color={rgb,255:red,255; green,0; blue,0}, mark=square*] coordinates{
		(0.1, 0.926880809)
		(0.2, 0.876775007)
		(0.3, 0.832983066)
		(0.4, 0.798307766)
		(0.5, 0.763353427)
		(0.6, 0.734184884)
		(0.7, 0.71185513)
		(0.8, 0.689265723)
		(0.9, 0.669562803)
		(1, 0.652884765)
	};
	\addlegendentry{Repeat layers}
	\addplot[color={rgb,255:red,0; green,255; blue,0}, mark=triangle*] coordinates{
		(0.1, 0.820279177)
		(0.2, 0.730358185)
		(0.3, 0.645675125)
		(0.4, 0.598171345)
		(0.5, 0.545900431)
		(0.6, 0.508586525)
		(0.7, 0.483546498)
		(0.8, 0.461980946)
		(0.9, 0.444676395)
		(1, 0.431564183)
	};
	\addlegendentry{U-sampling layers}
	\end{axis}
	\end{tikzpicture}
	\caption{Average CCA distance between the downsampling, repeating and upsampling portions of the trained network and its random initial states (trained on the VCTK dataset).}
	\label{fig:scenario_1_0}
\end{figure}
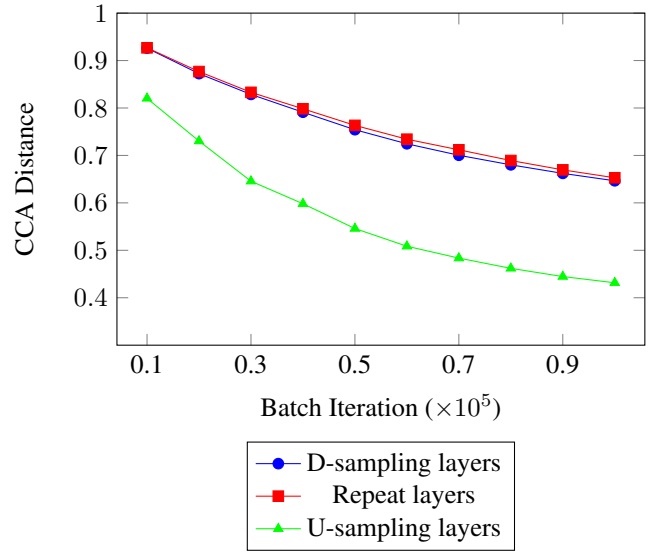
\begin{figure*}
	\begin{tikzpicture}
	\begin{axis}[
	width=.85\linewidth,
	height=5cm,
	bar width=0.15cm,
	ybar,
	enlarge x limits={abs=0.6cm},
	ylabel=CCA Distance,
	xlabel=Layer,
	xlabel style={name=xlabel},
	ymin=0.2, ymax=1,
	ytick={0.2, 0.3, 0.4, 0.5, 0.6, 0.7, 0.8, 0.9, 1},
	symbolic x coords={D1, D2, D3, DC, R1, R2, R3, R4, R5, R6, R7, R8, R9, UC, U1, U2, Out},
	xtick={D1, D2, D3, DC, R1, R2, R3, R4, R5, R6, R7, R8, R9, UC, U1, U2, Out},
	legend style={at={(1.2,0.72)}},
	]
	\addplot[fill={rgb,255:red,0; green,0; blue,0}] coordinates{
		(D1, 0.995265128)
		(D2, 0.939658381)
		(D3, 0.881979824)
		(DC, 0.887858838)
		(R1, 0.950785723)
		(R2, 0.934243442)
		(R3, 0.933316858)
		(R4, 0.934012932)
		(R5, 0.932909889)
		(R6, 0.929816413)
		(R7, 0.924464046)
		(R8, 0.913033926)
		(R9, 0.889344057)
		(UC, 0.97543542)
		(U1, 0.725691842)
		(U2, 0.810155133)
		(Out, 0.769834312)
	};
	\addlegendentry{$0.1\times10^{5}$ BIs}
	\addplot[fill={rgb,255:red,255; green,255; blue,255}, postaction={pattern=north east lines}] coordinates{
		(D1, 0.975263429)
		(D2, 0.7598371499)
		(D3, 0.620149263)
		(DC, 0.661550225)
		(R1, 0.831259943)
		(R2, 0.776563334)
		(R3, 0.77885172)
		(R4, 0.77724971)
		(R5, 0.776130969)
		(R6, 0.768551171)
		(R7, 0.753553577)
		(R8, 0.728904834)
		(R9, 0.679115585)
		(UC, 0.8951591088)
		(U1, 0.365983422)
		(U2, 0.440978749)
		(Out, 0.481480445)
	};
	\addlegendentry{$0.5\times10^{5}$ BIs}
	\addplot[fill={rgb,255:red,255; green,255; blue,255}] coordinates{
		(D1, 0.95081901)
		(D2, 0.64401135)
		(D3, 0.466416249)
		(DC, 0.524992852)
		(R1, 0.7203537862)
		(R2, 0.665458838)
		(R3, 0.673149512)
		(R4, 0.672189948)
		(R5, 0.669999936)
		(R6, 0.659725778)
		(R7, 0.641958305)
		(R8, 0.615165762)
		(R9, 0.557961023)
		(UC, 0.797768338)
		(U1, 0.274934408)
		(U2, 0.294947739)
		(Out, 0.358606247)
	};
	\addlegendentry{$1\times10^{5}$ BIs}
	\end{axis}
	\end{tikzpicture}
	\caption{CCA distance between each layer of a network at different stages of training and its random initial states, where \enquote{BI} denotes batch iteration trained on the VCTK dataset.}
	\label{fig:scenario_1_1}
\end{figure*}
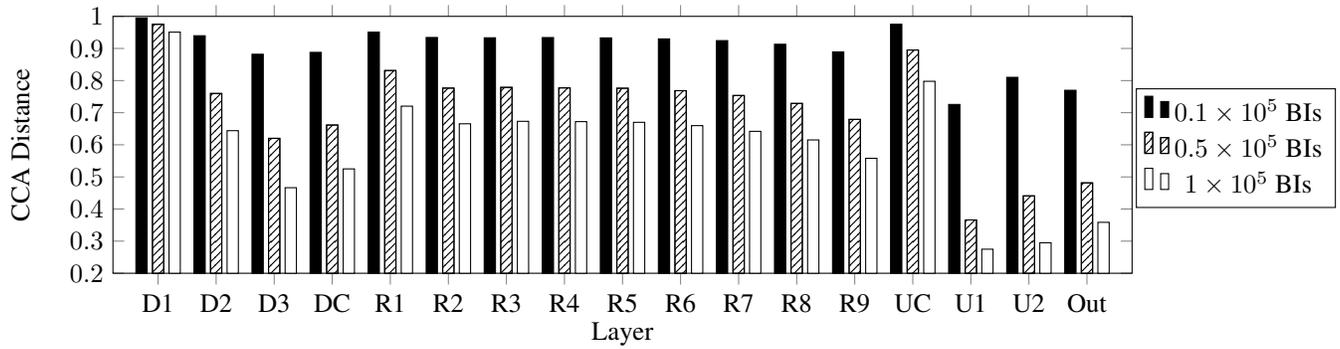
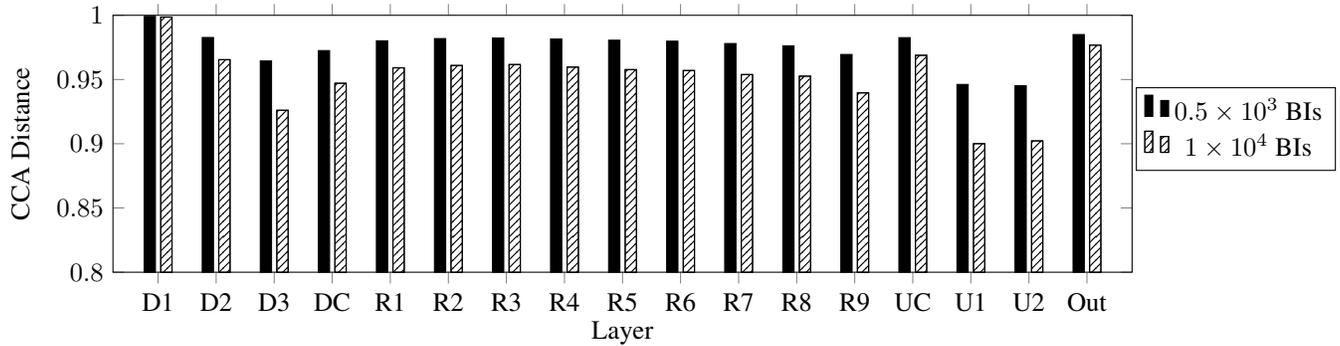
\begin{figure*}
	\begin{tikzpicture}
	\begin{axis}[
	width=.85\linewidth,
	height=5cm,
	bar width=0.15cm,
	ybar,
	enlarge x limits={abs=0.6cm},
	ylabel=CCA Distance,
	xlabel=Layer,
	xlabel style={name=xlabel},
	ymin=0.8, ymax=1,
	ytick={0.8, 0.85, 0.9, 0.95, 1},
	symbolic x coords={D1, D2, D3, DC, R1, R2, R3, R4, R5, R6, R7, R8, R9, UC, U1, U2, Out},
	xtick={D1, D2, D3, DC, R1, R2, R3, R4, R5, R6, R7, R8, R9, UC, U1, U2, Out},
	legend style={at={(1.2,0.72)}},
	]
	\addplot[fill={rgb,255:red,0; green,0; blue,0}] coordinates{
		(D1, 0.999181348)
		(D2, 0.982712308)
		(D3, 0.96444064)
		(DC, 0.9725259918)
		(R1, 0.9801484)
		(R2, 0.981867988)
		(R3, 0.982278736)
		(R4, 0.981518595)
		(R5, 0.980578394)
		(R6, 0.979829054)
		(R7, 0.978006906)
		(R8, 0.976221288)
		(R9, 0.969502039)
		(UC, 0.982595082)
		(U1, 0.946171326)
		(U2, 0.945198093)
		(Out, 0.985001413)
	};
	\addlegendentry{$0.5\times10^{3}$ BIs}
	\addplot[fill={rgb,255:red,255; green,255; blue,255}, postaction={pattern=north east lines}] coordinates{
		(D1, 0.998560423)
		(D2, 0.965440136)
		(D3, 0.926063088)
		(DC, 0.94708691)
		(R1, 0.959090542)
		(R2, 0.960926354)
		(R3, 0.961660374)
		(R4, 0.959653957)
		(R5, 0.957714566)
		(R6, 0.957001629)
		(R7, 0.95387059)
		(R8, 0.952601758)
		(R9, 0.9396032695)
		(UC, 0.968845872)
		(U1, 0.9000577)
		(U2, 0.9022234959)
		(Out, 0.976740572)
	};
	\addlegendentry{$1\times10^{4}$ BIs}
	\end{axis}
	\end{tikzpicture}
	\caption{CCA distance between each layer of a network at different stages of training during transfer learning from the initial states of the previous optimally trained network. Transfer learning was conducted using the VCC2018 dataset from a network previously trained on the VCTK dataset. \enquote{BI} denotes the number of batch iterations trained.}
	\label{fig:scenario_2_1}
\end{figure*}
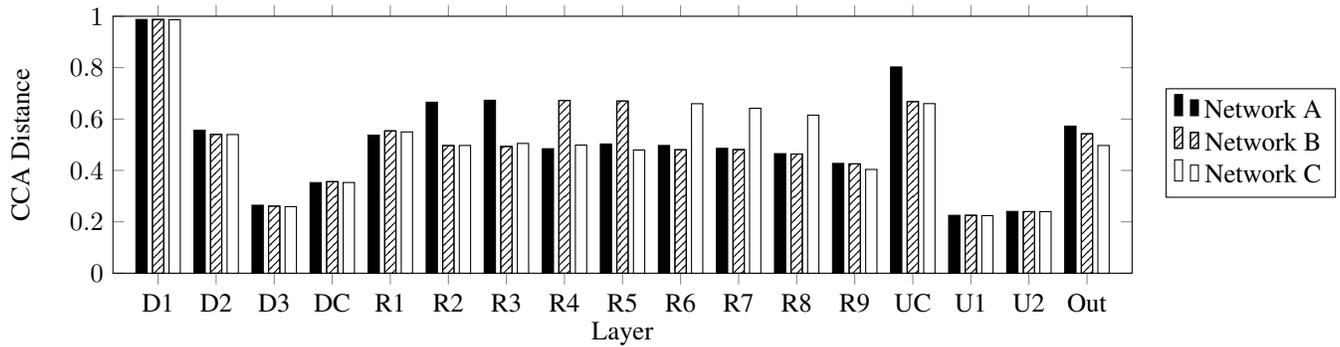
\begin{figure*}
	\begin{tikzpicture}
	\begin{axis}[
	width=.85\linewidth,
	height=5cm,
	bar width=0.15cm,
	ybar,
	enlarge x limits={abs=0.6cm},
	ylabel=CCA Distance,
	xlabel=Layer,
	xlabel style={name=xlabel},
	ymin=0, ymax=1,
	ytick={0, 0.2, 0.4, 0.6, 0.8, 1},
	symbolic x coords={D1, D2, D3, DC, R1, R2, R3, R4, R5, R6, R7, R8, R9, UC, U1, U2, Out},
	xtick={D1, D2, D3, DC, R1, R2, R3, R4, R5, R6, R7, R8, R9, UC, U1, U2, Out},
	legend style={at={(1.2,0.72)}},
	]
	\addplot[fill={rgb,255:red,0; green,0; blue,0}] coordinates{
		(D1, 0.987643822)
		(D2, 0.556232956)
		(D3, 0.264650408)
		(DC, 0.352641735)
		(R1, 0.537359323)
		(R2, 0.665458838)
		(R3, 0.673149512)
		(R4, 0.484596464)
		(R5, 0.502986193)
		(R6, 0.497140801)
		(R7, 0.485941412)
		(R8, 0.465408904)
		(R9, 0.427380499)
		(UC, 0.802745472)
		(U1, 0.225269318)
		(U2, 0.240546095)
		(Out, 0.572485638)
	};
	\addlegendentry{Network A}
	\addplot[fill={rgb,255:red,255; green,255; blue,255}, postaction={pattern=north east lines}] coordinates{
		(D1, 0.987690467)
		(D2, 0.540203789)
		(D3, 0.260945982)
		(DC, 0.356396171)
		(R1, 0.553991848)
		(R2, 0.496705031)
		(R3, 0.493215706)
		(R4, 0.672189948)
		(R5, 0.669999936)
		(R6, 0.480595446)
		(R7, 0.481029435)
		(R8, 0.463803264)
		(R9, 0.42569137)
		(UC, 0.667986926)
		(U1, 0.225441034)
		(U2, 0.239801686)
		(Out, 0.542665421)
	};
	\addlegendentry{Network B}
	\addplot[fill={rgb,255:red,255; green,255; blue,255}] coordinates{
		(D1, 0.986551122)
		(D2, 0.539693091)
		(D3, 0.259153817)
		(DC, 0.353092204)
		(R1, 0.549900915)
		(R2, 0.497177328)
		(R3, 0.505114426)
		(R4, 0.498670042)
		(R5, 0.479486367)
		(R6, 0.659725778)
		(R7, 0.641958305)
		(R8, 0.615165762)
		(R9, 0.403915505)
		(UC, 0.66011289)
		(U1, 0.224053699)
		(U2, 0.239741869)
		(Out, 0.497102076)
	};
	\addlegendentry{Network C}
	\end{axis}
	\end{tikzpicture}
	\caption{CCA distance between each layer of three different optimally trained networks with varying frozen layers and an optimally trained network with no frozen layers. Network A was trained with the parameters of layers R2 and R3 frozen, network B was trained with the parameters of layers R4 and R5 frozen, and network C was trained with the parameters of layers R6, R7 and R8 frozen. All three networks are extremely similar in terms of their acoustic output when compared with the same optimally trained network with no layers frozen.}
	\label{fig:scenario_3_1}
\end{figure*}

\textbf{Experiment 1}: To assess how close the optimally trained network's learned representations were to their random initialisations, SVCCA was used to compare networks at 0 and their optimal number of batch iterations. The number of optimal batch iterations for networks trained on the VCTK and VCC2018 datasets were found to be approximately $1\times10^{5}$ and $0.9\times10^{5}$, respectively.

Figures \ref{fig:scenario_1_0} and \ref{fig:scenario_1_1} show the CCA distance between the learned representations of layers in the network at different stages of training and their random initialisation. Both figures show a greater correlation of similarity in the learned network representations of the repeating 1D CNN layers (R1-R9) and their random initial states when compared to the less similar downsampling and upsampling portions of the network. Both figures show networks trained using the VCTK dataset however, similar results were observed with the VCC2018 dataset.

The extreme similarity observed at D1 can be seen as a fundamental trait of these networks. During pre-experiments, this trait was also seen in training StarGAN-VC \cite{Kameoka2018}. We removed the GLU of the first downsampling layer to check if this was preventing the first convolution from learning as much as it could.  However, the same trait was still observed.

The results show that the learned network representations of these optimally trained networks remain close to their initial random states, especially in the repeating 1D CNN layers, which are reportedly responsible for the main conversion process \cite{Kaneko2019cycle}. 

\textbf{Experiment 2}: The optimal model trained on the VCTK dataset in experiment 1 was used as the initial state for training during transfer learning on the VCC2018 dataset. Figure \ref{fig:scenario_2_1} shows that the learned parameters of the entire network remained extremely close to its initial network representations (the parameters learned from training on the VCTK dataset). The model converged after approximately $6000$ batch iterations and suffered from partial modal collapse.

The similarity between target reference and converted synthesised samples  transfer learning model was poor when compared with the original model trained on the same dataset. This could be due to the difference in speaker regions across datasets.

\textbf{Experiment 3}: The random initial state of the models trained in experiment 1 were used to train networks with various frozen layers in the repeating portion of the network. A total of three networks were evaluated with various frozen layers, the first of which froze R2 and R3. The second network froze R4 and R5, and the third network froze R6, R7 and R8. Figure \ref{fig:scenario_3_1} shows the similarity of these networks when compared against the optimally trained model from experiment 1. The repeating 1D layers again showed a high degree of similarity in their learned network representations. All networks were extremely similar in terms of their acoustic output when compared with the optimally trained model.

\textbf{Experiment 4}: The random initial state of the models trained in experiment 1 were used to train networks with differing numbers of repeating 1D layers. Six models were trained with 3, 5, 7, 11, 13 and 15 repeating layers in addition to the previously trained model from experiment 1, which had 9 repeating layers. It was observed that, in general, the audio quality of the models using 3, 5, 7 and 9 repeating layers sounded better than the models using 11, 13 and 15 layers. However, each model included at least one instance of having a worse quality of output than their counterparts for various different source-target pairs.

It was also observed that, as the number of repeating layers increased, the modification of speaker identity was more pronounced. In other words, the output from models with a greater number of repeating layers had clearer accents than the output of those with fewer repeating layers. However, at some points the modification of speaker identity was so pronounced that the intelligibility of the audio began to deteriorate. Also, as the amount of repeating layers of the network increased, so the overall level of noise increased. Networks consisting of 3 and 5 repeating layers struggled to converge whilst networks using 11, 13 and 15 suffered from a vanishing gradient.

\section{Conclusions} \label{sec:conclusion}

In the research reported here, we provide new insights into the interpretability of Generative Adversarial Networks (GANs) for Voice Conversion (VC). Using a network architecture based on StarGAN-VC2 \cite{Kaneko2019star}, we conducted an investigation into the learned representations of the network over a range of different experimental conditions. The results showed that there is at least one local optimum that lies close to the random initial states of the network. It was also found that it is the number of repeating layers in the network architecture that has a noticeable effect on the quality of the output speech.  In general, as the number of repeating layers in the network increased, so too did the noise and certain aspects of speaker identity became more pronounced. Future work will involve looking more into the importance of network depth in GANs for VC.

\bibliographystyle{IEEEbib}
\bibliography{refs}

\begin{thebibliography}{10}

\bibitem{Wang}
Xintao Wang, Ke~Yu, Shixiang Wu, Jinjin Gu, Yihao Liu, Chao Dong, Yu~Qiao, and
  Chen Change~Loy,
\newblock ``Esrgan: Enhanced super-resolution generative adversarial
  networks,''
\newblock in {\em Proceedings of the European Conference on Computer Vision
  (ECCV)}, 2018, pp. 0--0.

\bibitem{Zhu2017}
Jun-Yan Zhu, Taesung Park, Phillip Isola, and Alexei~A Efros,
\newblock ``Unpaired image-to-image translation using cycle-consistent
  adversarial networks,''
\newblock in {\em Proceedings of the IEEE international conference on computer
  vision}, 2017, pp. 2223--2232.

\bibitem{Kaneko2019cycle}
Takuhiro Kaneko, Hirokazu Kameoka, Kou Tanaka, and Nobukatsu Hojo,
\newblock ``Cyclegan-vc2: Improved cyclegan-based non-parallel voice
  conversion,''
\newblock in {\em ICASSP 2019-2019 IEEE International Conference on Acoustics,
  Speech and Signal Processing (ICASSP)}. IEEE, 2019, pp. 6820--6824.

\bibitem{Zhang2018}
Quan-shi Zhang and Song-Chun Zhu,
\newblock ``Visual interpretability for deep learning: a survey,''
\newblock {\em Frontiers of Information Technology \& Electronic Engineering},
  vol. 19, no. 1, pp. 27--39, 2018.

\bibitem{Mohammadi2017}
Seyed~Hamidreza Mohammadi and Alexander Kain,
\newblock ``An overview of voice conversion systems,''
\newblock {\em Speech Communication}, vol. 88, pp. 65--82, 2017.

\bibitem{Toda2016}
Tomoki Toda, Ling-Hui Chen, Daisuke Saito, Fernando Villavicencio, Mirjam
  Wester, Zhizheng Wu, and Junichi Yamagishi,
\newblock ``The voice conversion challenge 2016,''
\newblock in {\em Interspeech}, 2016, pp. 1632--1636.

\bibitem{Lorenzo-Trueba2018}
Jaime Lorenzo-Trueba, Junichi Yamagishi, Tomoki Toda, Daisuke Saito, Fernando
  Villavicencio, Tomi Kinnunen, and Zhenhua Ling,
\newblock ``The voice conversion challenge 2018: Promoting development of
  parallel and nonparallel methods,''
\newblock {\em arXiv preprint arXiv:1804.04262}, 2018.

\bibitem{Stylianou1998}
Yannis Stylianou, Olivier Capp{\'e}, and Eric Moulines,
\newblock ``Continuous probabilistic transform for voice conversion,''
\newblock {\em IEEE Transactions on speech and audio processing}, vol. 6, no.
  2, pp. 131--142, 1998.

\bibitem{Kawanami2003}
Hiromichi Kawanami, Yohei Iwami, Tomoki Toda, Hiroshi Saruwatari, and Kiyohiro
  Shikano,
\newblock ``Gmm-based voice conversion applied to emotional speech synthesis,''
\newblock in {\em Eighth European Conference on Speech Communication and
  Technology}, 2003.

\bibitem{Nakamura2012}
Keigo Nakamura, Tomoki Toda, Hiroshi Saruwatari, and Kiyohiro Shikano,
\newblock ``Speaking-aid systems using gmm-based voice conversion for
  electrolaryngeal speech,''
\newblock {\em Speech Communication}, vol. 54, no. 1, pp. 134--146, 2012.

\bibitem{Felps2009}
Daniel Felps, Heather Bortfeld, and Ricardo Gutierrez-Osuna,
\newblock ``Foreign accent conversion in computer assisted pronunciation
  training,''
\newblock {\em Speech communication}, vol. 51, no. 10, pp. 920--932, 2009.

\bibitem{Abe1990}
Masanobu Abe, Satoshi Nakamura, Kiyohiro Shikano, and Hisao Kuwabara,
\newblock ``Voice conversion through vector quantization,''
\newblock {\em Journal of the Acoustical Society of Japan (E)}, vol. 11, no. 2,
  pp. 71--76, 1990.

\bibitem{Toda2007}
Tomoki Toda, Alan~W Black, and Keiichi Tokuda,
\newblock ``Voice conversion based on maximum-likelihood estimation of spectral
  parameter trajectory,''
\newblock {\em IEEE Transactions on Audio, Speech, and Language Processing},
  vol. 15, no. 8, pp. 2222--2235, 2007.

\bibitem{Desai2009}
Srinivas Desai, E~Veera Raghavendra, B~Yegnanarayana, Alan~W Black, and Kishore
  Prahallad,
\newblock ``Voice conversion using artificial neural networks,''
\newblock in {\em 2009 IEEE International Conference on Acoustics, Speech and
  Signal Processing}. IEEE, 2009, pp. 3893--3896.

\bibitem{Mohammadi2014}
Seyed~Hamidreza Mohammadi and Alexander Kain,
\newblock ``Voice conversion using deep neural networks with
  speaker-independent pre-training,''
\newblock in {\em 2014 IEEE Spoken Language Technology Workshop (SLT)}. IEEE,
  2014, pp. 19--23.

\bibitem{Goodfellow}
Ian Goodfellow, Jean Pouget-Abadie, Mehdi Mirza, Bing Xu, David Warde-Farley,
  Sherjil Ozair, Aaron Courville, and Yoshua Bengio,
\newblock ``Generative adversarial nets,''
\newblock in {\em Advances in neural information processing systems}, 2014, pp.
  2672--2680.

\bibitem{KarrasNVIDIA}
Tero Karras, Samuli Laine, and Timo Aila,
\newblock ``A style-based generator architecture for generative adversarial
  networks,''
\newblock in {\em Proceedings of the IEEE Conference on Computer Vision and
  Pattern Recognition}, 2019, pp. 4401--4410.

\bibitem{Dong2018}
Hao-Wen Dong, Wen-Yi Hsiao, Li-Chia Yang, and Yi-Hsuan Yang,
\newblock ``Musegan: Multi-track sequential generative adversarial networks for
  symbolic music generation and accompaniment,''
\newblock in {\em Thirty-Second AAAI Conference on Artificial Intelligence},
  2018.

\bibitem{Hsu}
Chin-Cheng Hsu, Hsin-Te Hwang, Yi-Chiao Wu, Yu~Tsao, and Hsin-Min Wang,
\newblock ``Voice conversion from unaligned corpora using variational
  autoencoding wasserstein generative adversarial networks,''
\newblock {\em arXiv preprint arXiv:1704.00849}, 2017.

\bibitem{Kaneko2017}
Takuhiro Kaneko, Hirokazu Kameoka, Kaoru Hiramatsu, and Kunio Kashino,
\newblock ``Sequence-to-sequence voice conversion with similarity metric
  learned using generative adversarial networks.,''
\newblock in {\em INTERSPEECH}, 2017, vol. 2017, pp. 1283--1287.

\bibitem{Kaneko2019star}
Takuhiro Kaneko, Hirokazu Kameoka, Kou Tanaka, and Nobukatsu Hojo,
\newblock ``Stargan-vc2: Rethinking conditional methods for stargan-based voice
  conversion,''
\newblock {\em arXiv preprint arXiv:1907.12279}, 2019.

\bibitem{Paul2019}
Dipjyoti Paul, Yannis Pantazis, and Yannis Stylianou,
\newblock ``Non-parallel voice conversion using weighted generative adversarial
  networks,''
\newblock {\em Proc. Interspeech 2019}, pp. 659--663, 2019.

\bibitem{Lee}
Hung-Yi Lee and Yu~Tsao,
\newblock ``Generative adversarial network and its applications to speech
  processing and natural language processing,'' Speech Processing Laboratory,
  National Taiwan University.
  \url{http://speech.ee.ntu.edu.tw/~tlkagk/GAN_3hour.pdf},
\newblock (Accessed: November 19, 2019).

\bibitem{Dauphin2017}
Yann~N Dauphin, Angela Fan, Michael Auli, and David Grangier,
\newblock ``Language modeling with gated convolutional networks,''
\newblock in {\em Proceedings of the 34th International Conference on Machine
  Learning-Volume 70}. JMLR. org, 2017, pp. 933--941.

\bibitem{Taigman}
Yaniv Taigman, Adam Polyak, and Lior Wolf,
\newblock ``Unsupervised cross-domain image generation,''
\newblock {\em arXiv preprint arXiv:1611.02200}, 2016.

\bibitem{Choi2018}
Yunjey Choi, Minje Choi, Munyoung Kim, Jung-Woo Ha, Sunghun Kim, and Jaegul
  Choo,
\newblock ``Stargan: Unified generative adversarial networks for multi-domain
  image-to-image translation,''
\newblock in {\em Proceedings of the IEEE conference on computer vision and
  pattern recognition}, 2018, pp. 8789--8797.

\bibitem{Kameoka2018}
Hirokazu Kameoka, Takuhiro Kaneko, Kou Tanaka, and Nobukatsu Hojo,
\newblock ``Stargan-vc: Non-parallel many-to-many voice conversion using star
  generative adversarial networks,''
\newblock in {\em 2018 IEEE Spoken Language Technology Workshop (SLT)}. IEEE,
  2018, pp. 266--273.

\bibitem{voynov2019rpgan}
Andrey Voynov and Artem Babenko,
\newblock ``Rpgan: Gans interpretability via random routing,''
\newblock {\em arXiv preprint arXiv:1912.10920}, 2019.

\bibitem{mahendran2015understanding}
Aravindh Mahendran and Andrea Vedaldi,
\newblock ``Understanding deep image representations by inverting them,''
\newblock in {\em Proceedings of the IEEE conference on computer vision and
  pattern recognition}, 2015, pp. 5188--5196.

\bibitem{Raghu2017}
Maithra Raghu, Justin Gilmer, Jason Yosinski, and Jascha Sohl-Dickstein,
\newblock ``Svcca: Singular vector canonical correlation analysis for deep
  learning dynamics and interpretability,''
\newblock in {\em Advances in Neural Information Processing Systems}, 2017, pp.
  6076--6085.

\bibitem{Dumoulin2016}
Vincent Dumoulin, Jonathon Shlens, and Manjunath Kudlur,
\newblock ``A learned representation for artistic style,''
\newblock {\em arXiv preprint arXiv:1610.07629}, 2016.

\bibitem{huang2017arbitrary}
Xun Huang and Serge Belongie,
\newblock ``Arbitrary style transfer in real-time with adaptive instance
  normalization,''
\newblock in {\em Proceedings of the IEEE International Conference on Computer
  Vision}, 2017, pp. 1501--1510.

\bibitem{He2016}
Kaiming He, Xiangyu Zhang, Shaoqing Ren, and Jian Sun,
\newblock ``Deep residual learning for image recognition,''
\newblock in {\em Proceedings of the IEEE conference on computer vision and
  pattern recognition}, 2016, pp. 770--778.

\bibitem{Miyato}
Takeru Miyato and Masanori Koyama,
\newblock ``cgans with projection discriminator,''
\newblock {\em arXiv preprint arXiv:1802.05637}, 2018.

\bibitem{Long}
Jonathan Long, Evan Shelhamer, and Trevor Darrell,
\newblock ``Fully convolutional networks for semantic segmentation,''
\newblock in {\em Proceedings of the IEEE conference on computer vision and
  pattern recognition}, 2015, pp. 3431--3440.

\bibitem{Ulyanov}
Dmitry Ulyanov, Andrea Vedaldi, and Victor Lempitsky,
\newblock ``Instance normalization: The missing ingredient for fast
  stylization,''
\newblock {\em arXiv preprint arXiv:1607.08022}, 2016.

\bibitem{Hotelling1992}
Harold Hotelling,
\newblock ``Relations between two sets of variates,''
\newblock in {\em Breakthroughs in statistics}, pp. 162--190. Springer, 1992.

\bibitem{Hardoon2004}
David~R Hardoon, Sandor Szedmak, and John Shawe-Taylor,
\newblock ``Canonical correlation analysis: An overview with application to
  learning methods,''
\newblock {\em Neural computation}, vol. 16, no. 12, pp. 2639--2664, 2004.

\bibitem{Morcos2018}
Ari Morcos, Maithra Raghu, and Samy Bengio,
\newblock ``Insights on representational similarity in neural networks with
  canonical correlation,''
\newblock in {\em Advances in Neural Information Processing Systems}, 2018, pp.
  5727--5736.

\bibitem{Gautham2015}
Gautham~J Mysore,
\newblock ``Can we automatically transform speech recorded on common consumer
  devices in real-world environments into professional production quality
  speech?—a dataset, insights, and challenges,''
\newblock {\em IEEE Signal Processing Letters}, vol. 22, no. 8, pp. 1006--1010,
  2014.

\bibitem{Veaux2017}
Christophe Veaux, Junichi Yamagishi, and Kirsten MacDonald,
\newblock ``Cstr vctk corpus: English multi-speaker corpus for cstr voice
  cloning toolkit, [sound],'' Datashare, Edinburgh.
  \url{https://doi.org/10.7488/ds/1994},
\newblock (Accessed: December 16, 2019).

\bibitem{Morise2016}
Masanori Morise, Fumiya Yokomori, and Kenji Ozawa,
\newblock ``World: a vocoder-based high-quality speech synthesis system for
  real-time applications,''
\newblock {\em IEICE TRANSACTIONS on Information and Systems}, vol. 99, no. 7,
  pp. 1877--1884, 2016.

\bibitem{Kaneko2017filter}
Takuhiro Kaneko, Hirokazu Kameoka, Nobukatsu Hojo, Yusuke Ijima, Kaoru
  Hiramatsu, and Kunio Kashino,
\newblock ``Generative adversarial network-based postfilter for statistical
  parametric speech synthesis,''
\newblock in {\em 2017 IEEE International Conference on Acoustics, Speech and
  Signal Processing (ICASSP)}. IEEE, 2017, pp. 4910--4914.

\bibitem{Kaneko2017specto}
Takuhiro Kaneko, Shinji Takaki, Hirokazu Kameoka, and Junichi Yamagishi,
\newblock ``Generative adversarial network-based postfilter for stft
  spectrograms.,''
\newblock in {\em INTERSPEECH}, 2017, pp. 3389--3393.

\bibitem{Oord2016wavenet}
Aaron van~den Oord, Sander Dieleman, Heiga Zen, Karen Simonyan, Oriol Vinyals,
  Alex Graves, Nal Kalchbrenner, Andrew Senior, and Koray Kavukcuoglu,
\newblock ``Wavenet: A generative model for raw audio,''
\newblock {\em arXiv preprint arXiv:1609.03499}, 2016.

\bibitem{Tamamori2017vocoder}
Akira Tamamori, Tomoki Hayashi, Kazuhiro Kobayashi, Kazuya Takeda, and Tomoki
  Toda,
\newblock ``Speaker-dependent wavenet vocoder.,''
\newblock in {\em Interspeech}, 2017, vol. 2017, pp. 1118--1122.

\bibitem{Liu}
Kun Liu, Jianping Zhang, and Yonghong Yan,
\newblock ``High quality voice conversion through phoneme-based linear mapping
  functions with straight for mandarin,''
\newblock in {\em Fourth International Conference on Fuzzy Systems and
  Knowledge Discovery (FSKD 2007)}. IEEE, 2007, vol.~4, pp. 410--414.

\bibitem{Kingma}
Diederik~P Kingma and Jimmy Ba,
\newblock ``Adam: A method for stochastic optimization,''
\newblock {\em arXiv preprint arXiv:1412.6980}, 2014.

\bibitem{mao2017least}
Xudong Mao, Qing Li, Haoran Xie, Raymond~YK Lau, Zhen Wang, and Stephen
  Paul~Smolley,
\newblock ``Least squares generative adversarial networks,''
\newblock in {\em Proceedings of the IEEE international conference on computer
  vision}, 2017, pp. 2794--2802.

\end{thebibliography}

\end{document}